\documentclass[3p]{elsarticle}
\usepackage{amsfonts}
\usepackage{amsmath}
\usepackage{amssymb}
\usepackage{graphicx}%

\biboptions{numbers,sort&compress}

\makeatletter
\def\ps@pprintTitle{%
 \let\@oddhead\@empty
 \let\@evenhead\@empty
 \def\@oddfoot{}%
 \let\@evenfoot\@oddfoot}
\makeatother
\begin{document}

\title{Quaternionic representation of the genetic code}

\author[ifly,utn]{C. Manuel Carlevaro}
\ead{manuel@iflysib.unlp.edu.ar}
\author[ifly,jaur]{Ramiro M. Irastorza}
\ead{rirastorza@iflysib.unlp.edu.ar}
\author[ifly,gamefi]{Fernando Vericat\corref{cor}}
\ead{vericat@iflysib.unlp.edu.ar}

\cortext[cor]{Corresponding author}
\address[ifly]{Instituto de F\'isica de L\'iquidos y Sistemas Biol\'ogicos, 59 Nro. 780, 1900, La Plata, Argentina.}
\address[utn]{Universidad Tecnol\'ogica Nacional, Facultad Regional Buenos Aires, Mozart Nro. 2300, C14071VT, Buenos Aires, Argentina.}
\address[jaur]{Instituto de Ingenier\'ia y Agronom\'ia, Universidad Nacional Arturo Jauretche, 1888 Florencio Varela, Buenos Aires, Argentina.}
\address[gamefi]{Grupo de Aplicaciones Matem\'aticas y Estad\'isticas de la Facultad de Ingenier\'ia (GAMEFI), Universidad Nacional de La Plata, Calle 115 y 48, 1900 La Plata, Argentina.}
 
 \begin{abstract}
A heuristic diagram of the evolution of the standard genetic code is
presented. It incorporates, in a way that resembles the energy levels of an
atom, the physical notion of broken symmetry and it is consistent with
original ideas by Crick on the origin and evolution of the code\ as well as
with the chronological order of appearence of the amino acids along the
evolution as inferred from work that mixtures known experimental results with
theoretical speculations. Suggested by the diagram we propose a Hamilton
quaternions based mathematical representation of the code as it stands
now-a-days. The central object in the description is a codon function that
assigns to each amino acid an integer quaternion in such a way that the
observed code degeneration is preserved. We emphasize the advantages of a
quaternionic representation of amino acids taking as an example the folding of
proteins. With this aim we propose an algorithm to go from the quaternions
sequence to the protein three dimensional structure which can be compared with
the corresponding experimental one stored at the Protein Data Bank. In our
criterion the mathematical representation of the genetic code in terms of
quaternions merits to be taken into account because it describes not only most
of the known properties of the genetic code but also opens new perspectives
that are mainly derived from the close relationship between quaternions and rotations.

\bigskip\bigskip


\end{abstract}
\maketitle

\section{Introduction}

The standard genetic code\cite{Crick1}, say the correspondence between the
sequence of \ nucleotide bases of mRNA molecules and the sequence of amino
acids in the ribosomal protein synthesis as occurring at the cells of most
of the animals and plants, is now-a-days fairly well known. The mRNA bases
belong to the set $\left\{ A,C,G,U\right\} $ where $A$ stands for adenine, $%
C $ for cytosine, $G$ for guanine and $U$ for uracil. Non-overlapping
triplets of consecutive bases (codons) encode just one of the $20$ standard
amino acids (see Appendix A) or a stop signal each one. In principle, there
is no any kind of separation between adjacent codons in the sequence. Of the 
$4^{3}=64$ possible different codons, $61$ translate into amino acids and
the remaining three determine a stop signal. We are then speaking about a
code of four letters that can form $64$ words three letters each. \ The
words translate into amino acids or the stop signal.

The mechanism that performs this translation involves a very sophisticated
molecular machinery which is no completely known yet. However, Crick%
\'{}%
s \ adaptor hypothesis\cite{Crick2} and further refinements\cite{Ibba1,Ibba2} are, in general, widely accepted as accurate enough as to describe,
at molecular level, the complex translation procedure in most of the cases.
The image currently accepted is that tRNA molecules act as intermediaries
(adaptors) between the template (mRNA) and the amino acids that will form
the protein. The amino acid to be incorporated into the protein chain is
covalently bonded to the tRNA $3%
{\acute{}}%
$ extreme (forming an aminoacyl-tRNA complex) at the time that, in another
part of the tRNA chain, a triplet of nucleotide bases (anticodon)
specifically interacts with the codon of the mRNA template that codifies the
amino acid in question. The bases of the anticodon are just the
complementary ones of the corresponding codon bases (read in the direction $5%
{\acute{}}%
\rightarrow3%
{\acute{}}%
$) and the interactions manifest as hydrogen bonds between complementary
bases.

Skipping over for the moment the molecular details of the translation and
restricting ourselves to the correspondence \textit{codons}$\rightarrow $%
\textit{amino acids} in itself, we reproduce in Figure 1 a classical
presentation of the standard genetic code. The structure of the code is
evident. Each codon codifies just one amino acid or (in the case of the
codons $UAA$, $UAG$ and $UGA$) the stop signal. The code is degenerate in
the sense that, except for the amino acids methionine (met) and tryptophan
(trp) that are codified by a single codon each one, all the other amino
acids are codified by two or more codons.

\begin{figure}[h!]
\centering
\includegraphics[width=1.0\textwidth]{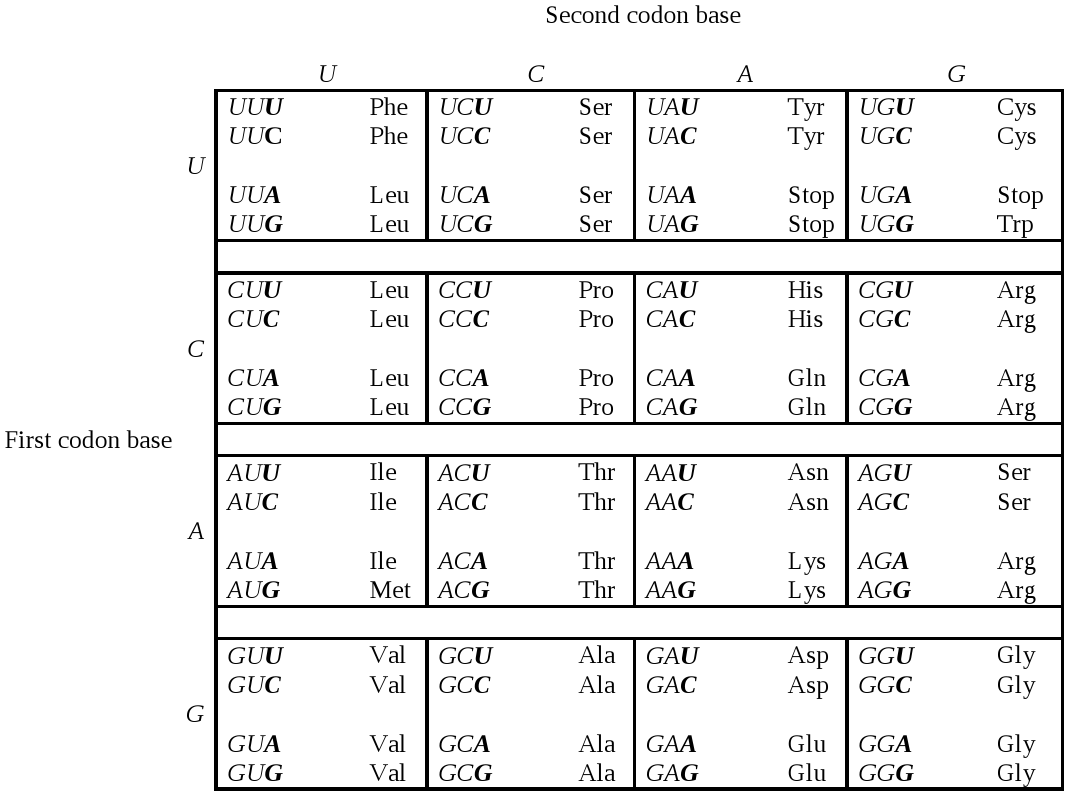} 
\caption{Text book picture of the standard genetic code. The three letters
convention for the amino acids is used (see Appendix A) and the third base
in the codons is remarked in bold. The order of the codons is in the
direction $5^{\prime}\rightarrow 3^{\prime}$. The codon \textit{AUG} besides
to codify the amino acid methionine (met)\textbf{\ }also determines the
starting point within the mRNA sequence for the protein synthesis.}
\label{fig:1}
\end{figure}

One interesting related question that has received some attention is the
origin and evolution of the genetic code. The proposals in this direction
are obviously rather speculative\cite{Jukes1,Wong1,Osawa1,Hartman1,Sanchez1}. However, Crick%
\'{}%
s scenario\cite{Crick3} according to which originally only a few amino acids
were coded by most of the possible three bases codons and that, in
subsequent steps, some of those codons were substituting the amino acid they
coded by a new one until eventually the code became frozen in its present
form, seems reasonable and very attractive. In particular, the idea of an
increasing number of amino acids to be coded, can be correlated with the
studies on the evolution of the amino acids abundance\cite{Miller1,Trifonov1}.

A step further in relation with the genetic code includes several efforts
done in order to give mathematical models for describing the present
structure of the code and how it has evolved in order to reach this state%
\cite{Gonzalez1,Hornos1,Sciarrino1}. The main mathematical tools are tensor
algebras and group theory. In particular, in Ref. \cite{Hornos1} the authors
use the physical concept of broken symmetry to find a mathematical group
with a 16-dimensional representation (the highly degenerate primitive code)
which can be written as the product of simpler groups that describe the
pattern of redundancies observed in Figure 1. The approach gives a very
elegant physical explanation of the code degeneration. However, perhaps
because it concerns the application of a relatively complicated mathematical
tool to a subject dominated by researchers with main formation in
disciplines other than Mathematics and Physics, the work has been taken just
as a valuable exercise in classification\cite{Maddox1,Stewart1} .

In this work we propose a mathematical description of the genetic code too,
but it is based on a tool that, in our judgement, is very friendly and, at
the same time, very powerful as to open new perspectives beyond of simply
giving a representation of the code structure. We are talking about the
Hamilton quaternions\cite{Hamilton1,Hamilton2}. These mathematical
objects are a sort of generalization of the complex numbers and obey an
algebra in many aspects similar to theirs but with the very important (for
our purposes) property that the product is, in general, non commutative (see
Appendix B). In addition, the quaternions are ideal for representing
rotations with important advantages over the classical matrix
representation. This fact has of course already been recognized by
bioinformaticians in writing routines involving the tertiary structure of
proteins. We must mention that Petoukhov has also applied quaternions to
descrbe the genetic code but from a very different point of view\cite{Petoukhov1}.

Our journey starts by presenting in the next Section a diagram for the
evolution of the genetic code that incorporates the concept of broken
symmetry in a way that resembles the energy levels of an atom. Actually, our
interest is in the present form of the code, however the evolution diagram
gives a picture of the correspondence \textit{bases triplets}$\rightarrow $%
\textit{amino acids} that will help us with the mathematical representation
of this correspondence by means of quaternions. Moreover, despite the high
degree of speculation that exists in any model for the origin and evolution
of the genetic code, we can give to our diagram an interpretation which is
consistent with the above mentioned ideas by Crick on the subject\cite{Crick3}. Thus, inspired by this diagram, in Section III we proceed to
represent the relationship between the codons and amino acids by using
quaternions. First we assign an integer quaternion (Lipschitz integer) to
each one of the four nucleotide bases and then, suggested by the diagram
structure, we consider a codons function that gives as result the
assignation of a quaternion to each one of the amino acids. The explicit
form of this function involves simple quaternionic operations (products and
sums) that automatically accounts for the degeneration of amino acids
encoded by four, three or two codons and includes, in addition to the
quaternions assigned to the four bases, an extra number of quaternions,
related with the splitting of the "atomic levels" due to the symmetry
breaking during the evolution, which, in principle, are indeterminate. These
\ extra quaternions are determined by demanding that the degeneration for
amino acids encoded by more than four (concretely six) codons be also
verified. In order that this scheme works in practice we need to explicitly
give the four quaternions for the bases. Of the infinitely many options the
one we choose clearly has a Pythagorean flavor: we consider a subset of four
quaternions from the complete set of eight prime integer quaternions with
norm $7$. The subset we take does not contain pairs of conjugate
quaternions, four being the maximum cardinality for a subset with this
property\cite{Davidoff1}. Once\ a quaternion of this subset has been
assigned to each of the four nucleotide bases, the quaternion corresponding
to each amino acid is directly determined by the above mentioned function.
This way the quaternionic description of the genetic code is completed. In
order to remark the potentiality of the quaternionic representation of amino
acids for opening new perspectives beyond the description of the genetic
code degeneration, we appeal to another fundamental question: the protein
folding problem, say the establishment of the native tertiary structure of
the protein from the knowledge of its amino acids sequence (primary
structure)\cite{Creighton1,BenNaim1}. The protein folding problem is 
\textit{per se} a phenomenal task that in some sense can be considered as
experimentally solved through X-ray diffraction, Nuclear Magnetic Resonance
and other techniques. However, theoretically the problem remains unsolved
and a lot of work has been done by many researchers since the middle of the
past century in order to develop a computational procedure that allows
predicting the tertiary structure of a given protein from its amino acids
sequence. Here we avoid to mention the lot of methods proposed to attack the
question and simply give our own, maybe rather heuristic, approach as to
show the advantages of associating amino acids with quaternions. This will
be done in Section IV were we show the procedure\ that we have designed in
order to go from the amino acids quaternions to the coordinates of the
backbone alpha-carbon atoms of a protein whose spatial structure we assume
is the native one for the given amino acids sequence. These coordinates can
be compared with those experimentally obtained as given in the Protein Data
Bank (PDB)\cite{PDB}. The procedure involves a set of real quaternions
associated with the order of the amino-acids in the chain so that each
amino-acid in a protein is represented by an integer quaternion (type
quaternion) and a real quaternion (order quaternion). If this quaternions
are the same ones for all the proteins, then the protein folding problem
would be solved. In this work we limit ourselves to show how the type and
order quaternions can be used to transform the primary structure of a given
protein into its spatial configuration. The problem of obtaining the set of
order quaternions which is adequate to all proteins (if it exists), say the
possibility of solving the protein folding problem, is left for future work.

Two Appendices, one with the one and three letters convention for
identifying the 20 standard amino acids and another one with the main
properties of the quaternions are finally given for completeness.

\section{A diagram for the evolution of the genetic code}

In Figure 2 we show the diagram that we propose to take account of the
evolution of the genetic code. It is mainly inspired in pioneering ideas by
Crick\cite{Crick3} and also in the physical concept of broken symmetry,
first applied in relation with the genetic code by Hornos and Hornos\cite%
{Hornos1}. 
\begin{figure}[tbp]
\centering
\includegraphics[width=0.9\textwidth]{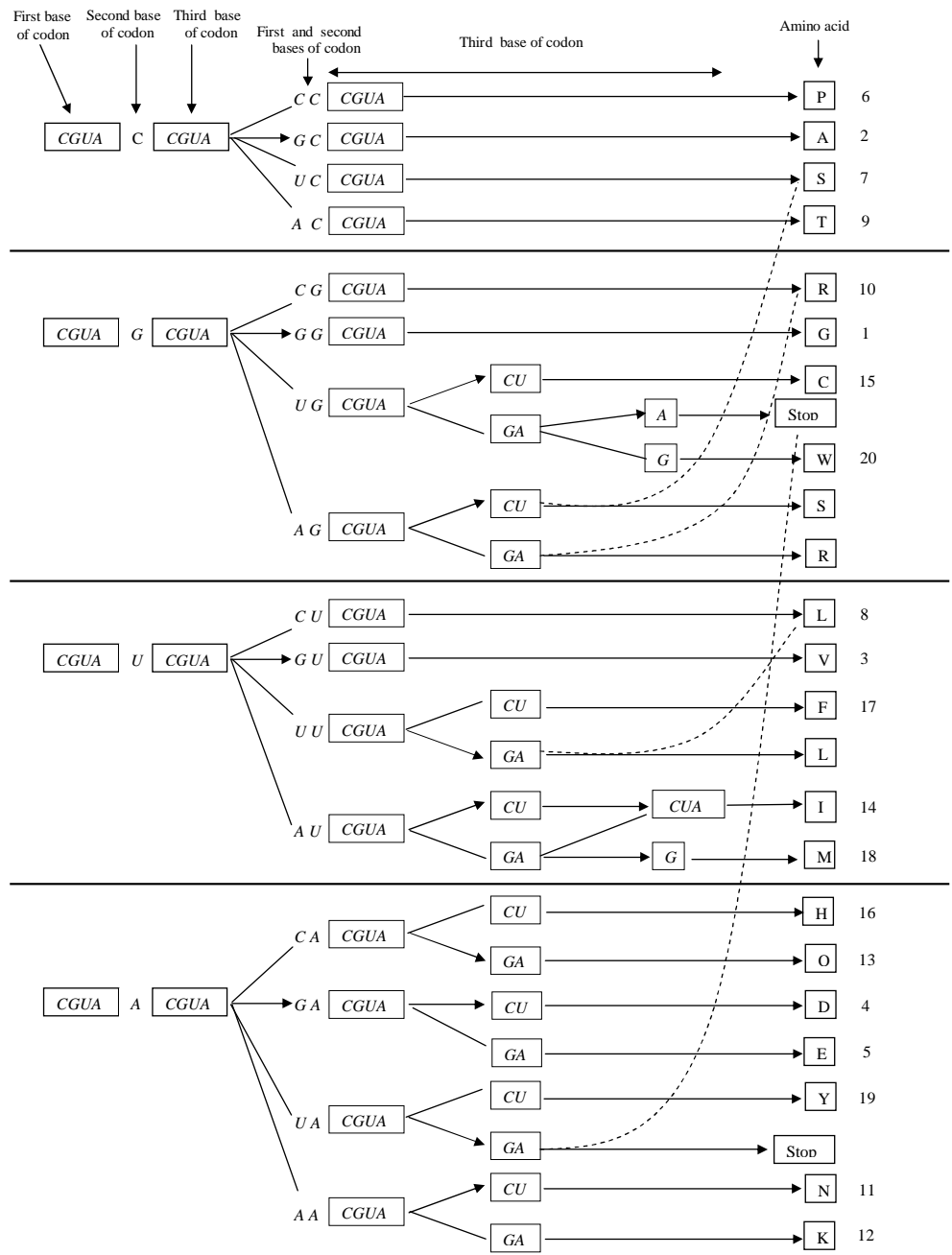} 
\caption{Authors proposal for the genetic code evolution. The one letter
convention for amino acids is used (see Appendix A). The direction of the
temporal evolution is from left to right. Rectangles with two or more bases
implies degeneration with respect to those ones. The broken lines link
different sets of codons that encode the same amino acid in the case of
sixfold degeneration. Arrows and common lines indicate what codons follow
codifying the same amino acid and what will start to codify a new one,
respectively, after the symmetry is broken (see text). The natural numbers
at the right side of the diagram give the temporal order of the amino acids
in the Trifonov consensus scale\protect\cite{Trifonov1}.}
\label{fig:2}
\end{figure}

According to Crick if the genetic code is at present time a triplet code, in
the sense that the reading mechanism moves along three bases at each step,
then it must always have been a triplet code since otherwise a loss of
Darwinian fitness can occur. Thus we assume that the codons were always
formed by three bases of the set $\left\{ A,C,G,U\right\} $. We must mention
that Crick also have analyzed the plausibility of primitive nucleic acids
constituted by just two bases. However even if this were the case, since the
passage to a four bases system had to occur in some moment of the evolution
without to substantially alter the\ message carried by the old two bases
chain (Principle of continuity), we can take the four bases alphabet as
being always available since a given moment at the origins of the code.
Therefore we accept that since the beginning codons are triplets of bases
chosen from the set $\left\{ A,C,G,U\right\} $. Moreover, we consider that,
in the first evolution steps, only the second base of the codon was
effective in codifying amino acids. Accordingly only four amino acids could
be codified, each one by one of the four bases $C$, $G$, $U$ and $A$
independently of which the first and third bases are. In the diagram this
fact is denoted with a rectangle containing the four letters. This is
consistent with Crick\'{}s suggestion that only a few amino acids were coded
at the beginning. According with the diagram, $C$ would codify alanine (A); $%
G$, glycine (G); $U$, valine (V) and $A$ aspartic acid (D) whatever the
first and third bases are. \ It is worth noting here that the four amino
acids that we assume were the first ones to be codified are the first four
in the Trifonov\cite{Trifonov1} consensus temporal order scale for the
appearance of the amino acids (column of natural numbers in Figure 2). The
four amino acids A, G, V and D were also the first four that appeared under
simulation of the primitive earth conditions in Miller experiments\cite%
{Miller1}.

As the left part of diagram shows, our version of the primitive code is
highly degenerate: in principle each of the four amino acids, A,G,V and D,
could be encoded by $4^{2}=16$ codons. Physically the idea of degeneration
is closely related with the concept of symmetry and a very illustrative form
to think about these concepts is by doing an analogy with the energy levels
of an atom. In our case we would have four levels indexed each one with the
letter corresponding to the second codon base, say $C$, $G$, $U$ and $A$
(main quantum number). We thus assume that, as the code evolves, the
symmetry that causes that the amino acid codification be independent of the
first base of the codon, disappears for some reason. The reason could be
that with time the recognition mechanism becomes more precise as to
differentiate between two codons with distinct first base. \ Because of this
symmetry breaking, a part of the degeneration also disappears. In the
diagram each of the four initial levels splits into four new levels, one for
each of the possible bases ($C$, $G$, $U$ and $A$) at the first place of the
codon (secondary quantum number). Now we have a total of $16$ levels indexed
each one by two letters (the first and second bases of the codon). Each
level is fourfold degenerate in the codons third base. One of the new levels
follows codifying the same amino acid as before that the level splits
whereas the other three codify a new amino acid each. We indicate with an
arrow the four groups of codons that conserve the amino acid and with a
simple line those that substitute the amino acid by a new one. Note that the
codons that follow codifying the same amino acid are those whose first base
is guanine ($G$). This is consistent with the above mentioned temporal order
of appearance and with the present time correct assignation of amino acids
in the case of fourfold degeneration as is shown in Figure 1. This way $9$
new amino acids (that with the old four sum $13$) and the stop signal are
coded. Note also that we assume that the amino acids serine (S) and leucine
(L) at that moment were codified by two groups of codons: S by $UC$(\textit{%
third base arbitrary}) and $AG$(\textit{third base arbitrary}), whereas L by 
$CU$(\textit{third base arbitrary}) and $UU$(\textit{third base arbitrary}).

As the code follows evolving it suffers new breaking of symmetry so that the
third base of some codons bring into use or, in the atomic analogy, some of
the fourfold degenerate levels split into two levels\ each one twofold
degenerate. Those levels pointed out with an arrow follow codifying the same
amino acid whereas the other levels substitute it for a new one. Eventually,
in subsequent steps, a few of the twofold degenerate levels split once more
given two non-degenerate levels each. This is the case of codons that codify
methionine (M), tryptophan (W) and (again) the stop signal. The case of
isoleucine (I) is a particular one since the split level coincides with the
twofold one which represents the two codons that follow codifying the same
amino acid. This way, isoleucine is the only amino acid which is coded by
three codons. The stop signal is also threefold degenerate since it is coded
by two groups of codons one twofold degenerate and the other one
non-degenerate. At this step of the evolution the code frozen to give its
present form. It is worth mentioning that the code evolution gives as a
particular result that the amino acids serine (S), arginine (R) and leucine
(L) are at present coded by two groups of codons each one. In the three
cases one of the groups is fourfold degenerate and the other one is twofold
degenerate, so that these amino acids are the only three which are sixfold
degenerates. We point out this property in the diagram with a broken line
linking the two groups of codons. The two groups of codons that codify the
stop signal are also linked by a broken line.

\section{\protect\bigskip Mathematical representation of the genetic code}

We proceed now to describe the genetic code by using quaternions. Define the
sets:

\begin{equation}
\mathcal{B}=\left\{ C,G,U,A\right\} ,  \tag{1}  \label{1}
\end{equation}

\begin{equation}
\mathcal{A}=\left\{ \text{P,A,S,T,R,G,C,W,L,V,F,I,M,H,Q,D,E,Y,N,K,Stop}%
\right\}  \tag{2}  \label{2}
\end{equation}
and

\begin{equation}
\mathbf{H}_{7,\text{ red.}}\left( 
\mathbb{Z}
\right) =\left\{ \left( 2,1,1,1\right) ,\text{ }\left( 2,-1,1,1\right) ,%
\text{ }\left( 2,1,-1,1\right) ,\text{ }\left( 2,1,1,-1\right) \right\} . 
\tag{3}  \label{3}
\end{equation}%
We propose a quaternionic representation of the genetic code according to
the following scheme:

\begin{equation}
\begin{array}{ccc}
\mathcal{B}^{3} & \longrightarrow & \mathcal{A} \\ 
\downarrow &  & \downarrow \\ 
\mathbf{H}_{7,\text{ red.}}^{3}\left( 
\mathbb{Z}
\right) & \longrightarrow & \mathbf{H}\left( 
\mathbb{Z}
\right)%
\end{array}
\tag{4}  \label{4}
\end{equation}
where $\mathbf{H}\left( 
\mathbb{Z}
\right) $ denotes the set of integer quaternions (see Appendix B). $\mathcal{%
B}^{3}$ is the set of the $64$ codons and we assume that the correspondence $%
\mathcal{B}^{3}\rightarrow\mathcal{A}$ is the present day standard genetic
code as described by Figure 1, whereas the function $\mathcal{B}%
^{3}\rightarrow\mathbf{H}_{7,\text{ red.}}^{3}\left( 
\mathbb{Z}
\right) $ assigns to each codon a triplet of quaternions of the set $\mathbf{%
H}_{7,\text{ red.}}\left( 
\mathbb{Z}
\right) $ (the subindex red. is for reduced). This set is a maximum
cardinality subset of 
\begin{equation*}
\mathbf{H}_{7}\left( 
\mathbb{Z}
\right) =\left\{ \left( a_{0},a_{1},a_{2},a_{3}\right)
:a_{0},a_{1},a_{2},a_{3}\in%
\mathbb{Z}
\text{; }a_{0}^{2}+a_{1}^{2}+a_{2}^{2}+a_{3}^{2}=7\text{, }a_{0}>0\text{
and\ even}\right\}
\end{equation*}
with the property that it does not contain pairs of conjugate quaternions.
The set $\mathbf{H}_{7}\left( 
\mathbb{Z}
\right) $ has $7+1=8$ elements\cite{Davidoff1} and so $\mathbf{H}_{7,\text{
red.}}\left( 
\mathbb{Z}
\right) $ has $4$ quaternions as it should be. It is worth-noting that all
the integer quaternions in $\mathbf{H}_{7}\left( 
\mathbb{Z}
\right) $ are prime quaternions in the sense that they can not be expressed
as the product of two integer quaternions if neither of them can be the unit
quaternion $\left( 1,0,0,0\right) $. This is consistent with the fact that
an integer quaternion is prime if and only if its norm is a prime number\cite%
{Davidoff1}. Note that taking the nucleotide bases as prime quaternions
gives them a certain character of \textit{elemental }molecules. Apart from
this, the election of $\mathbf{H}_{7,\text{ red.}}\left( 
\mathbb{Z}
\right) $ may seem rather arbitrary. However we are just looking for a
quaternionic representation of the genetic code so that, whatever the set of
quaternions that we assign to the codons is, the important issue is that the
function $\mathbf{H}_{7,\text{ red.}}^{3}\left( 
\mathbb{Z}
\right) $ $\rightarrow$ $\mathbf{H}\left( 
\mathbb{Z}
\right) $ preserves the essential properties of the correspondence $\mathcal{%
B}^{3}\rightarrow$ $\mathcal{A}$.

In what follows, in order to simplify the notation, we assign natural
numbers to identify the bases and the amino acids: $C\rightarrow1$, $%
G\rightarrow2$, $U\rightarrow3$, $A\rightarrow4$ and P$\rightarrow1$, A$%
\rightarrow2$, S$\rightarrow3$, T$\rightarrow4$, R$\rightarrow5$, G$%
\rightarrow6$, C$\rightarrow7$, W$\rightarrow8$, L$\rightarrow9$, V$%
\rightarrow10$, F$\rightarrow11$, I$\rightarrow12$, M$\rightarrow13$, H$%
\rightarrow14$, Q$\rightarrow15$, D$\rightarrow16$, E$\rightarrow17$, Y$%
\rightarrow18$, N$\rightarrow19$, K$\rightarrow20$, Stop$\rightarrow21$.

Inspired by the diagram of Figure 2 we define the quaternionic function

\begin{equation}
\ \ \ \ \ 
\begin{array}{c}
F:\mathbf{H}_{7,\text{ red.}}^{3}\left( 
\mathbb{Z}
\right) \rightarrow\mathbf{H}\left( 
\mathbb{Z}
\right) \text{ \ \ \ \ \ \ \ \ \ \ \ \ \ \ \ \ \ \ } \\ 
\text{ \ \ }\left( q_{\beta},q_{\gamma},q_{\delta}\right) \rightarrow
\alpha_{i}=F\left[ \left( q_{\beta},q_{\gamma},q_{\delta}\right) \right]%
\end{array}
\tag{5}  \label{5}
\end{equation}
by (see Appendix B for the operations between quaternions):

\begin{align}
\begin{split}
\text{ P}\rightarrow\alpha_{1}&=q_{1}q_{1} \qquad \qquad\qquad \quad \;
(\beta=1\text{, }\gamma=1,\delta=1,2,3,4) \\
\text{A}\rightarrow\alpha_{2}&=q_{2}q_{1} \qquad \qquad\qquad \quad \;
(\beta=2,\gamma=1,\delta=1,2,3,4) \\
\text{S}\rightarrow\alpha_{3}&=q_{3}q_{1}=q_{4}q_{2}+\gamma_{2;13} \quad
(\beta=3,\gamma=1,\delta=1,2,3,4 \text{ or }\beta=4,\gamma=2,\delta=1,3) \\
\text{T}\rightarrow\alpha_{4}&=q_{4}q_{1} \qquad \qquad\qquad \quad \;
(\beta=4,\gamma=1,\delta=1,2,3,4) \\
\text{R}\rightarrow\alpha_{5}&=q_{1}q_{2}=q_{4}q_{2}+\gamma_{2;24} \quad
(\beta=1,\gamma=2,\delta=1,2,3,4 \text{ or }\beta=4,\gamma=2,\delta=2,4) \\
\text{G}\rightarrow\alpha_{6}&=q_{2}q_{2} \qquad \qquad\qquad \quad \;
(\beta=2,\gamma=2,\delta=1,2,3,4) \\
\text{C}\rightarrow\alpha_{7}&=q_{3}q_{2}+\gamma_{2;13} \qquad \qquad
(\beta=3,\gamma=2,\delta=1,3) \\
\text{W}\rightarrow\alpha_{8}&=q_{3}q_{2}+\gamma_{2;24}+\delta _{2;2} \quad
\; (\beta=3,\gamma=2,\delta=2) \\
\text{L}\rightarrow\alpha_{9}&=q_{1}q_{3}=q_{3}q_{3}+\gamma_{3;24} \quad
(\beta=1,\gamma=3,\delta=1,2,3,4 \text{ or }\beta=3,\gamma=3,\delta=2,4) \\
\text{V}\rightarrow\alpha_{10}&=q_{2}q_{3} \qquad \qquad\qquad \quad \;
(\beta=2,\gamma=3,\delta=1,2,3,4) \\
\text{F}\rightarrow\alpha_{11}&=q_{3}q_{3}+\gamma_{3;13} \qquad
\qquad(\beta=3,\gamma=3,\delta=1,3) \\
\text{I}\rightarrow\alpha_{12}&=q_{4}q_{3}+\gamma_{3;13}=q_{4}q_{3}+%
\gamma_{3;24}+\delta_{3;4} \qquad (\beta=4,\gamma=3,\delta=1,3,4) \\
\text{M}\rightarrow\alpha_{13}&=q_{4}q_{3}+\gamma_{3;24}+\delta_{3;2} \quad
\; (\beta=4,\gamma=3,\delta=2)) \\
\text{H}\rightarrow\alpha_{14}&=q_{1}q_{4}+\gamma_{4;13} \qquad \qquad
(\beta=1,\gamma=4,\delta=1,3) \\
\text{Q}\rightarrow\alpha_{15}&=q_{1}q_{4}+\gamma_{4;24} \qquad
\qquad(\beta=1,\gamma=4,\delta=2,4) \\
\text{D}\rightarrow\alpha_{16}&=q_{2}q_{4}+\gamma_{4;13} \qquad
\qquad(\beta=2,\gamma=4,\delta=1,3) \\
\text{E}\rightarrow\alpha_{17}&=q_{2}q_{4}+\gamma_{4;24} \qquad
\qquad(\beta=2,\gamma=4,\delta=2,4) \\
\text{Y}\rightarrow\alpha_{18}&=q_{3}q_{4}+\gamma_{4;13} \qquad
\qquad(\beta=3,\gamma=4,\delta=1,3) \\
\text{N}\rightarrow\alpha_{19}&=q_{4}q_{4}+\gamma_{4;13} \qquad
\qquad(\beta=4,\gamma=4,\delta=1,3)) \\
\text{K}\rightarrow\alpha_{20}&=q_{4}q_{4}+\gamma_{4;24} \qquad
\qquad(\beta=4,\gamma=4,\delta=2,4)) \\
\text{Stop}\rightarrow\alpha_{21}&=q_{3}q_{2}+\gamma_{2;24}+%
\delta_{2;4}=q_{3}q_{4}+\gamma_{4;24} \qquad (\beta=3,\gamma=2,\delta=4 
\text{ or } \gamma=4,\delta=2,4)
\end{split}
\tag{6}  \label{6}
\end{align}

>From these expressions we can appreciate the importance of working with
objects that obey a non commutative algebra. In fact, if the quaternions
product where commutative then amino acids A and R would have associated the
same quaternion and the same would occur with S and L.

In Eq.(\ref{6}), the quaternions $\gamma_{\text{i;jk}}$ accounts for the
level splitting when the second base of codon is i and the third base is jk$%
= $ $13$ ($CU$) or $24$ ($GA$). Analogously, the quaternion $\delta_{\text{%
i:j}}$ accounts for the level splitting when the second base of the codon is
i and the third base is j$=2$ ($G$) or $4$\ ($A$). Thus, in principle we
have as unknown quaternions $\gamma_{2;13}$, $\gamma_{2;24}$, $\gamma_{3;13}$%
, $\gamma_{3;24}$, $\gamma_{4;13}$, $\gamma_{4;24}$ and $\delta_{2;2}$, $%
\delta_{2;4}$, $\delta_{3;2}$ and $\delta_{3;4}$.

Of the $10$ \ unknown quaternions we can find $5$, say $\gamma_{2;13}$, $%
\gamma_{2;24}$, $\gamma_{3;13}$, $\gamma_{3;24}$, $\gamma_{4;24}$, by
requiring that those amino acids which are coded by two different groups of
codons (case of codons sixfold degenerates or codons that codify the stop
signal) have associated an unique quaternion and also that the two ways to
reach isoleucine (I)\ give the same quaternion (see Figure 2), so we must
solve the system

\begin{equation}
\left\{ 
\begin{array}{l}
q_{3}q_{1}=q_{4}q_{2}+\gamma_{2;13}\text{ \ \ \ \ \ \ \ \ \ \ \ \ \ \ \ \ \
\ \ \ \ \ \ }\left( \alpha_{3}\right) \\ 
q_{1}q_{2}=q_{4}q_{2}+\gamma_{2;24}\text{ \ \ \ \ \ \ \ \ \ \ \ \ \ \ \ \ \
\ \ \ \ \ \ }\left( \alpha_{5}\right) \\ 
q_{1}q_{3}=q_{3}q_{3}+\gamma_{3;24}\text{\ \ \ \ \ \ \ \ \ \ \ \ \ \ \ \ \ \
\ \ \ \ \ \ }\left( \alpha _{9}\right) \\ 
q_{4}q_{3}+\gamma_{3;13}=q_{4}q_{3}+\gamma_{3;24}+\delta_{3;4}\text{ \ \ \ \
\ }\left( \alpha_{12}\right) \\ 
q_{3}q_{2}+\gamma_{2;24}+\delta_{2;4}=q_{3}q_{4}+\gamma_{4;24}\text{ \ \ \ \
\ }\left( \alpha_{21}\right) .%
\end{array}
\right.  \tag{7}  \label{7}
\end{equation}
The solution is:

\begin{equation}
\begin{array}{l}
\gamma_{2;13}=q_{3}q_{1}-q_{4}q_{2}\text{ \ \ \ \ \ \ \ \ \ \ \ \ \ \ \ \ \
\ \ \ \ \ \ \ } \\ 
\gamma_{2;24}=q_{1}q_{2}-q_{4}q_{2}\text{ \ \ \ \ \ \ \ \ \ \ \ \ \ \ \ \ \
\ \ \ \ \ \ \ } \\ 
\gamma_{3;13}=q_{1}q_{3}-q_{3}q_{3}+\delta_{3;4}\text{ \ \ \ \ \ \ \ \ \ \ \
\ \ \ \ \ \ } \\ 
\gamma_{3;24}=q_{1}q_{3}-q_{3}q_{3}\text{ \ \ \ \ \ \ \ \ \ \ \ \ \ \ \ \ \
\ \ \ \ \ \ } \\ 
\gamma_{4;24}=q_{3}q_{2}+q_{1}q_{2}-q_{4}q_{2}-q_{3}q_{4}+\delta_{2;4}.\text{
}%
\end{array}
\tag{8}  \label{8}
\end{equation}
To obtain the quaternions $\delta_{2;2}$, $\delta_{2;4}$, $\delta_{3;2}$ and 
$\delta_{3;4}$ we assign to those levels that can not split more (non
degenerate levels) the product of the quaternions associated with each of
the corresponding bases: $\alpha_{8}=q_{3}q_{2}q_{2}$; $%
\alpha_{13}=q_{4}q_{3}q_{2}$; $\alpha_{21}=q_{3}q_{2}q_{4}$; $%
\alpha_{12}=q_{4}q_{3}q_{4}$. This way we have

\begin{align}
\delta_{2;2} & =q_{3}q_{2}q_{2}-q_{3}q_{2}-\gamma_{2;24}  \notag \\
\delta_{3;2} & =q_{4}q_{3}q_{2}-q_{4}q_{3}-\gamma_{3;24}  \notag \\
\delta_{2;4} & =q_{3}q_{2}q_{4}-q_{3}q_{2}-\gamma_{2;24}  \tag{9}  \label{9}
\\
\delta_{3;4} & =q_{4}q_{3}q_{4}-q_{4}q_{3}-\gamma_{3;24.}  \notag
\end{align}
\medskip\bigskip Finally for the remaining unknown quaternion $\gamma_{4;13}$
we propose:%
\begin{equation}
\gamma_{4;13}=-\gamma_{4;24}.  \tag{10}  \label{10}
\end{equation}

Eqs.(\ref{6}), (\ref{8}), (\ref{9}) and (\ref{10}) solve completely the
problem of assigning quaternions to the amino acids in such a way that the
pattern of redundancy of the genetic code is verified. Taking: $q_{1}=\left(
2,1,1,1\right) $, $q_{2}=\left( 2,-1,1,1\right) $, $q_{3}=\left(
2,1,-1,1\right) $ and $q_{4}=\left( 2,1,1,-1\right) $, we explicitly obtain

\begin{equation}
\begin{array}{lll}
\alpha_{1}=\left( 1,4,4,4\right) \text{\ \ \ } & \alpha_{8}=\left(
6,-15,-1,9\right) \text{\ \ \ \ \ \ } & \alpha_{15}=\left( 16,-3,7,1\right) 
\text{\ \ \ \ \ } \\ 
\alpha_{2}=\left( 3,0,6,2\right) \text{\ \ \ } & \alpha_{9}=\left(
3,6,0,2\right) \text{\ \ \ \ \ \ \ \ \ \ \ \ } & \alpha_{16}=\left(
-8,3,3,-3\right) \text{\ \ \ \ } \\ 
\alpha_{3}=\left( 3,2,0,6\right) \text{\ \ \ } & \alpha_{10}=\left(
5,2,2,4\right) \text{\ \ \ \ \ \ \ \ \ \ \ \ } & \alpha_{17}=\left(
18,-7,5,-1\right) \text{\ \ } \\ 
\alpha_{4}=\left( 3,6,2,0\right) \text{\ \ \ } & \alpha_{11}=\left(
2,17,1,3\right) \text{\ \ \ \ \ \ \ \ \ \ } & \alpha_{18}=\left(
-8,9,1,1\right) \text{\ \ \ \ \ \ } \\ 
\alpha_{5}=\left( 3,0,2,6\right) \text{\ \ \ } & \alpha_{12}=\left(
6,17,3,-3\right) \text{\ \ \ \ \ \ \ } & \alpha_{19}=\left(
-12,9,3,-5\right) \text{\ } \\ 
\alpha_{6}=\left( 1,-4,4,4\right) \text{\ } & \alpha_{13}=\left(
18,3,-1,3\right) \text{\ \ \ \ \ \ \ \ } & \alpha_{20}=\left(
14,-1,5,-3\right) \text{\ } \\ 
\alpha_{7}=\left( 3,-2,-6,8\right) \text{\ } & \alpha_{14}=\left(
-10,7,5,-1\right) \text{\ \ \ \ \ } & \alpha_{21}=\left( 18,-1,3,3\right) 
\text{\ }.\text{ }%
\end{array}
\tag{11}  \label{11}
\end{equation}
We will denote $\mathbf{H}_{\alpha}\left( 
\mathbb{Z}
\right) $ the set of quaternions assigned to the amino acids as given by Eq.(%
\ref{11}).

At first sight this set of quaternions could seem to say nothing special by
itself, however when we watch it more carefully we start to discover some
patterns of regularity or symmetries. The first thing that we observe is
that the norm of all these quaternions is odd: $N\left( \alpha_{i}\right)
=a_{0}^{2}+a_{1}^{2}+a_{2}^{2}+a_{3}^{2}\equiv1$ mod$\left( 2\right) $ $%
\left( i=1,2,\cdots,21\right) $ and can roughly be taken as a measure of the
information needed to codify the corresponding amino acid in the sense that
the larger the norm the larger the necessary information. In fact, taking
into account the multiplicative property of the quaternions norm we can
easily see from Eq. (\ref{6}) that those quaternions associated with amino
acids which need just the first and second codon bases to be recognized, say 
$\alpha
_{1},\alpha_{2},\alpha_{3},\alpha_{4},\alpha_{5},\alpha_{6},\alpha_{9}$ and $%
\alpha_{10}$, have as norm $N\left( \alpha_{i}\right) =N\left( q_{\beta
}q_{\gamma}\right) =N\left( q_{\beta}\right) N\left( q_{\gamma}\right) =49$
whereas those which need of the three bases to that effect, say the
quaternions $\alpha_{8}$ and $\alpha_{13}$ corresponding to the amino acids
methionine (M), tryptophan (W) and also $\alpha_{12}$ associated with the
amino acid isoleucine (I) and $\alpha_{21}$ with the stop signal, both
coming (in one of two possible ways) from a non degenerate level (see Figure
2 and also Eq. \ref{6}), have $N\left( \alpha_{i}\right) =N\left(
q_{\beta}q_{\gamma }q_{\delta}\right) =$ $N\left( q_{\beta}\right) N\left(
q_{\gamma}\right) N\left( q_{\delta}\right) =343$ . Here we have used the
fact that the norms of the quaternions that represent the nucleotide bases
are $N\left( q_{\beta }\right) =7$ ($\beta=1,2,3,4$). If the information
about what amino acids will be added during the protein synthesis is encoded
in the quaternions triplets $\left( q_{\beta},q_{\gamma},q_{\delta}\right) $
then for amino acids which are determined by quaternions of the type $%
\alpha_{i}=$ $q_{\beta }q_{\gamma}$ the lack of information is compensated
with the degeneration in the third base whereas for amino acids specified by
quaternions of the form $\alpha_{i}=q_{\beta}q_{\gamma}q_{\delta}$ there is
no lack of information and redundancy would be, in principle, not necessary.
The amino acids which are twofold degenerate have norms which lie, with just
one exception ($\alpha _{17}$), in between these extreme values.

We can also use the norm to divide the set $\mathbf{H}_{\alpha}\left( 
\mathbb{Z}
\right) $ into classes: the norm of the quaternions corresponding to four or
sixfold degenerate levels\textit{\ }verifies $N\left( \alpha_{i}\right)
\equiv1$ mod$\left( 4\right) $ whereas all the remaining quaternions, say $%
\alpha_{8},\alpha_{11},\alpha_{12},\alpha_{13},\alpha_{14},\alpha_{15},%
\alpha_{16},\alpha_{17},\alpha_{18},\alpha_{19},\alpha_{20}$ and $\alpha
_{21}$ that come from levels with lower degeneration, have norm that
fulfills $N\left( \alpha_{i}\right) \equiv3$ mod$\left( 4\right) $. The
exception is the quaternion corresponding to the amino-acid cysteine which
is coded by two codons but verifies $N\left( \alpha_{7}\right) =113\equiv1$
mod$\left( 4\right) $. At the respect we can say that in the \textit{%
euplotid nuclear }variant of the genetic code the codon \textit{UGA }%
codifies the amino acid C instead of the stop signal. If we consider this
variant then $\alpha_{7}$ would play in some sense the role of $\alpha_{21}$
and vice versa and the exception would be the stop signal which could be
eliminated of the discussion that mainly concerns with amino acids. However
since we are actually interested into the standard code we simply take the
quaternion $\alpha_{7}$ as the exception to the rule and momentarily ignore
it in our discussion here. The class of quaternions that verifies $N\left(
\alpha_{i}\right) \equiv3$ mod$\left( 4\right) $ can still be split into a
couple of groups: one ($\alpha_{15},\alpha_{16},\alpha_{18},\alpha_{19}$)
with $N\left( \alpha _{i}\right) \equiv3$ mod$\left( 8\right) $ and the
other one ($\alpha
_{8},\alpha_{11},\alpha_{12},\alpha_{13},\alpha_{14},\alpha_{17},\alpha
_{20},\alpha_{21}$) with $N\left( \alpha_{i}\right) \equiv7$ mod$\left(
8\right) $. Although we have not clear the actual meaning of this separation
we suspect that it has to do with symmetries involved in the translation
process at molecular level. Anyway we think that these simple observations
are enough as to give a preliminary idea about the potential usefulness of
quaternions to discover hidden patterns of symmetry inside the genetic code.

\section{Amino acids as quaternions and the folding of proteins}

As we have seen in the previous Section, our quaternionic representation of
the genetic code reproduces its structure, particularly the code redundancy
and allows to make evident some regularity patterns. However the point that
we wish to remark here is the special richness that gives to the description
the close relationship between quaternions and rotations (see Appendix B).
Because of the advantages of using quaternions to describe spatial
rotations, the association of amino acids with quaternions opens new
horizons beyond the genetic code representation. In this context, we
consider the suitability of this association to take account of the folding
of the proteins that the amino acids form.

The primary structure of a protein of $N$ amino acids is a sequence A$_{1}$,A%
$_{2}$,$\ldots $,A$_{N}$ with A$_{i}\in \mathcal{A}$ . The protein folding
problem consists in obtaining from this sequence the spatial coordinates of
each one of the atoms of all the amino acids that constitute the protein
when this one is in the native -or functional- state (tertiary structure).
As such we consider the one corresponding to the protein in physiological
solution whose coordinates can be obtained, after crystallization, by
application of, for example, X-ray diffraction methods. That is the case of
most of the proteins whose coordinates are stored at the PDB. In principle
we restrict ourselves to determine the coordinates for just the alpha-carbon
atoms of the chain which is not a severe restriction since it is known that
there exist very efficient algorithms for going from this trace
representation to the full atoms one\cite{Rotkiewicz1}. We also take into
account that, in our quaternionic representation, the amino acids sequence
is expressed as a sequence of quaternions $p_{1}$,$p_{2}$,$\ldots $,$p_{N}$
with $p_{i}\in \mathbf{H}_{\alpha }\left( 
\mathbb{Z}
\right) $. Under these conditions we proceed now to present an algorithm to
determine the spatial coordinates of the alpha-carbon atoms of the protein.

\begin{figure}[tbp]
\centering
\includegraphics[width=0.6\textwidth]{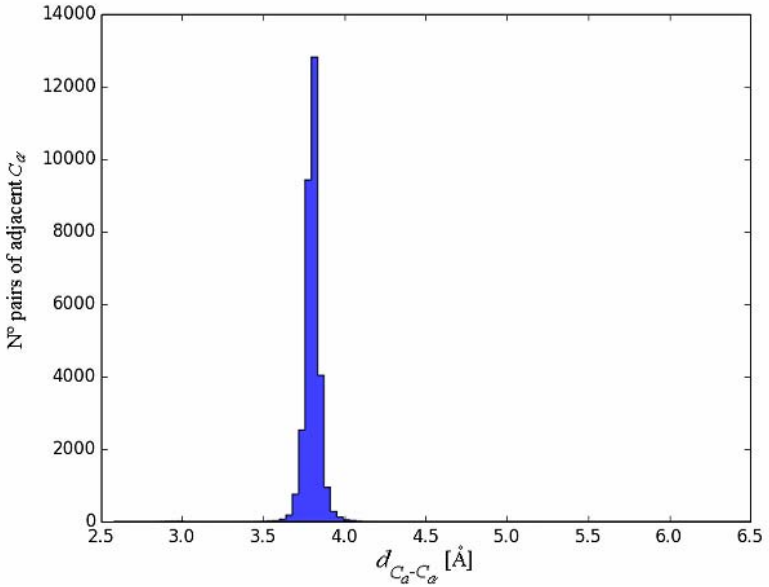} 
\caption{Histogram for the distance $d_{C_{\protect\alpha}-C_{\protect\alpha%
}}$ between adjacent alpha-carbon atoms. The distances were calculated from
the alpha-carbon atoms coordinates corresponding to a sample of $110$
proteins of different length stored at the PDB ($31332$ pairs of adjacent
alpha-carbon atoms). The mean value and the standard deviation are $%
\left\langle d_{C_{\protect\alpha}-C_{\protect\alpha}}\right\rangle =3.801$ $%
\mathring{A}$ and $\protect\sigma_{C_{\protect\alpha}-C_{\protect\alpha}}=%
\left[ \left\langle d_{C_{\protect\alpha}-C_{\protect\alpha%
}}^{2}\right\rangle -\left\langle d_{C_{\protect\alpha}-C_{\protect\alpha%
}}\right\rangle ^{2}\right] ^{1/2}=0.061$ \AA{}, respectively.}
\label{fig:3}
\end{figure}

First we observe that although adjacent alpha-carbon atoms are not
covalently bonded their distance is notably stable and take very similar
values for all the pairs within a given protein and also for those belonging
to different proteins, as the histogram of Figure 3 shows. So in our
calculations we assume that all these distances are equal to a unique value $%
d_{C_{\alpha }-C_{\alpha }}=3.80$ \AA{}. Thus we determine on the
unit sphere with center at the origin a point for each of the amino acids
(alpha-carbon atoms) in the protein sequence. To the last one we assign
directly the origin, the preceding one is located at the intersection
between the axis $z$ and the sphere surface (versor $\widehat{e}_{z}$). To
each of the remaining alpha-carbon atoms we assign a point on the sphere
surface that results of rotating the versor $\widehat{e}_{z}$ by a
quaternion (see Appendix B). For the $j$th alpha-carbon atom in the
sequence, the quaternion responsible of the rotation is denoted $\widehat{%
\beta }_{j\text{ }}$($j=1,2,\cdots ,N-2$). We then expand the chain of
alpha-carbon atoms from their location on the sphere into the back-bone
protein three dimensional configuration (See Figure 4) by means of the
following iterative procedure, where initially the$\ \overrightarrow{r}_{j}$%
\'{}%
s are on the sphere surface:

\ \ \ \ \ \ \ \ \ \ \ \ \ \ \ \ \ \ \ \ \ \ \ \ \ \ \ \ \ \ \ \ \ \ \ \ \ \
\ \ \ \ \ \ \ \ do $i=1,N-2$

$\ \ \ \ $\ $\ \ \ \ \ \ \ \ \ \ \ \ \ \ \ \ \ \ \ \ \ \ \ \ \ \ \ \ \ \ \ \
\ \ \ \ \ \ \ \ \ \ \ \ \ \delta \vec{r}=\vec{r}_{i+1}$

\ \ \ \ \ \ \ \ \ \ \ \ \ \ \ \ \ \ \ \ \ \ \ \ \ \ \ \ \ \ \ \ \ \ \ \ \ \
\ \ \ \ \ \ \ \ \ \ \ \ \ \ do $j=1,i$

$\ \ \ \ \ \ \ \ $\ $\ \ \ \ \ \ \ \ \ \ \ \ \ \ \ \ \ \ \ \ \ \ \ \ \ \ \ \
\ \ \ \ \ \ \ \ \ \ \ \ \ \ \ \ \ \ \vec {r}_{j}=\vec{r}_{j}+\delta\vec{r}$

\ \ \ \ \ \ \ \ \ \ \ \ \ \ \ \ \ \ \ \ \ \ \ \ \ \ \ \ \ \ \ \ \ \ \ \ \ \
\ \ \ \ \ \ \ \ \ \ \ \ \ \ end do

\ \ \ \ \ \ \ \ \ \ \ \ \ \ \ \ \ \ \ \ \ \ \ \ \ \ \ \ \ \ \ \ \ \ \ \ \ \
\ \ \ \ \ \ \ \ \ end do\ \ \ \ \ \ \ \ \ \ \ \ \ \ \ \ \ \ \ \ \ \ \ \ \ \
\ \ \ \ \ 

According to the algorithm, the distance between adjacent alpha-carbon atoms
is the unit so, to establish the correct distance, we must multiply the
final calculated coordinates by $d_{C_{\alpha }-C_{\alpha }}$.

\begin{figure}[tbp]
\centering
\includegraphics[width=0.5\textwidth]{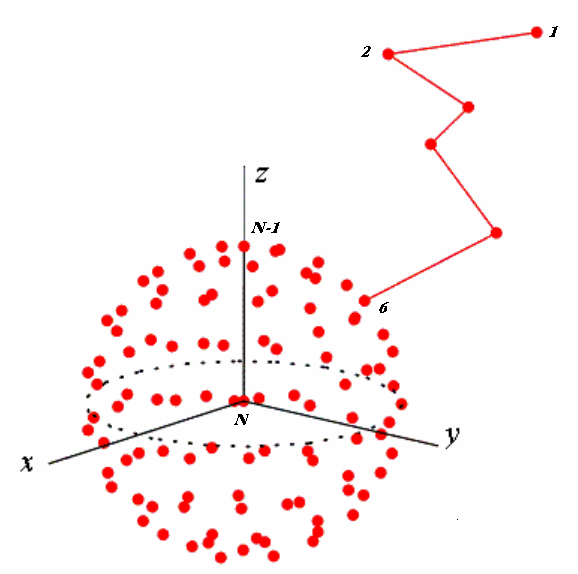} 
\caption{Development of the alpha-carbon atoms backbone of a hypothetical
protein of length $N$ from its position on the sphere surface into its
spatial configuration (schematic). The last two alpha-carbon atoms, as well
as some of the first ones, are labelled by their order number inside the
sequence.}
\label{fig:4}
\end{figure}

It remains to determine how to calculate the quaternions $\widehat{\beta }_{j%
\text{ }}$($j=1,2,\cdots ,N-2$). \ We do this in a somewhat heuristic way.
We take into account that the $j$th amino acid interacts in some way with
the $j-1$ previous amino acids in the sequence and also with the $N-j$
subsequent ones. Of course that in these interactions the effect of the
medium should be incorporated in some form, for example in the form of
effective interactions between amino acids. Actually we are trying for a
sort of decodification and so we are not directly interested into the
detailed form of the interactions, but we recognize that in any codification
of information that involves those interactions, some trace of their general
form should be. In general it is reasonable to think that the global
interaction includes two body, three body,..., until $N$ body (effective)
interactions so by analogy we choose with generality for \ $\widehat{\beta }%
_{j\text{ }}$ the normalized version of the quaternion\ 

\begin{equation}
\beta_{j}=\left( S_{j,1}^{<}+S_{j,2}^{<}+\cdots+S_{j,j-1}^{<}\right)
p_{j}+p_{j}\left( S_{j,1}^{>}+S_{j,2}^{>}+\cdots+S_{j,N-j}^{>}\right) 
\tag{12}  \label{12}
\end{equation}
with

\begin{equation}
S_{j,1}^{<}=\sum\limits_{1\leq r\leq j-1}c_{r}p_{r},\text{ \ \ \ }%
S_{j,2}^{<}=\sum\limits_{1\leq r<s\leq
j-1}c_{r}p_{r}c_{s}p_{s},\cdots,S_{j,j-1}^{<}=c_{1}p_{1}c_{2}p_{2}\cdots
c_{j-1}p_{j-1}  \tag{13}  \label{13}
\end{equation}
and

\begin{equation}
S_{j,1}^{>}=\sum\limits_{j+1\leq r\leq N}c_{r}p_{r},\ \ \
S_{j,2}^{>}=\sum\limits_{j+1\leq r<s\leq
N}c_{r}p_{r}c_{s}p_{s},\cdots,S_{j,N-j}^{>}=c_{j+1}p_{j+1}c_{j+2}p_{j+2}%
\cdots c_{N}p_{N},  \tag{14}  \label{14}
\end{equation}
where $c_{r}\in\mathbf{H}\left( 
\mathbb{R}
\right) $ ($r=1,2,\cdots,N$) are in principle unknown real quaternions to be
determined. It is worth mentioning that in our election of the form of Eq.(%
\ref{12}) we have taken into account the non commutativity of quaternions
too.

Even for proteins of length $N$ relatively small, the memory and computation
time required for evaluating the unknown quaternions $c_{1},c_{2},\cdots
,c_{N}$ using the complete expression given by Eq.(\ref{12}) for the $%
\beta_{j}$%
\'{}%
s are too large, at least for our computational facilities. Thus in the
calculations here we use the simplest version:

\begin{equation}
\beta_{j}=S_{j,1}^{<}p_{j}+p_{j}S_{j,1}^{>},  \tag{15}  \label{15}
\end{equation}
that, in our analogy, corresponds to consider just pair interactions in the
protein total potential energy.

Here we adjust the unknown quaternions by means of an optimization
technique. As such we use the particle swarm optimization (PSO) procedure of
Kennedy and Eberhart\cite{Kennedy1} taking as function of fitness the
difference between the coordinates of the alpha-carbon atoms calculated
following the previous procedure and the corresponding experimental ones as
read from the PDB. We take the rmsd (root-mean-square deviation) as a
measure of this difference, using to that effect Bosco K. Ho%
\'{}%
s implementation of Kabsch algorithm\cite{Kabsch1}. This way we assign to
each amino-acid in the primary structure of the protein, two quaternions: an
integer quaternion belonging to the set $\mathbf{H}_{\alpha}\left( 
\mathbb{Z}
\right) $ (\textit{type quaternion}) and a real one (\textit{order quaternion%
}) according to its position inside the protein chain.

\begin{figure}[tbp]
\centering
\includegraphics[width=0.5\textwidth]{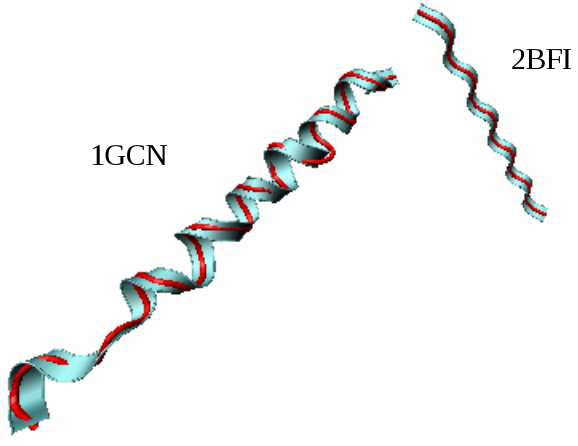} 
\caption{Trace representation of the alpha-carbon atoms backbone for the
small proteins 2BFI and 1GCN. Red (dark grey) tube: from the coordinates
obtained using our procedure. Cyan (light grey) ribbon: from the coordinates
stored at PDB.}
\label{fig:5}
\end{figure}

\begin{figure}[tbp]
\centering
\includegraphics[width=0.5\textwidth]{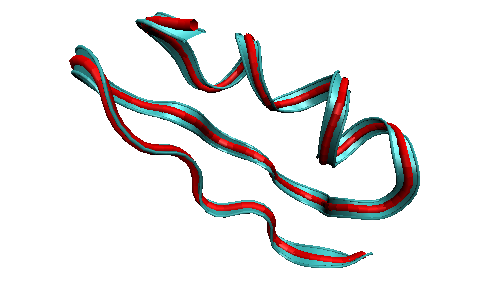} 
\caption{Trace representation of the alpha-carbon atoms backbone for the
protein 2CK5. Red (dark grey) tube: from the coordinates obtained using our
procedure. Cyan (light grey) ribbon: from the coordinates stored at PDB.}
\label{fig:6}
\end{figure}

In figures 5 to 8 we show the result of the application of our procedure to
five small peptides and proteins: in Figure 5 the synthetic peptide amyloid
fibril (PDB\ ID: 2BFI - length: $12$ amino acids) together with the hormone
glucagon (PDB\ ID: 1GCN - length: $29$ amino acids); in Figure 6 the ion
channel inhibitor osk1 toxin (PDB\ ID: 2CK5 - length: $31$ amino acids); in
Figue 7 a type III antifreeze protein (PDB ID:1HG7 - length: $66$ amino
acids); in figure 8 the "hydrogen atom" of proteins, say myoglobin (PDB\ ID:
1MBN - length: 153 amino acids). The two proteins of Figure 5 were adjusted
simultaneously so that the order quaternions for 2BFI are the same ones as
the first $12$ of 1GCN. The $29$ order quaternions we have obtained for 1GCN
differ of the first $29$ ones of the remainder proteins instead. With
respect to this last fact we must mention that, at least within the error
(rmsd) considered here, the set of quaternions we found for a given protein
by fitting the alpha-carbon atoms coordinates is not unique. This is an
important point since otherwise the possibility of finding a common set of
order quaternions valid for all the proteins would be definitively closed.
In the figures we compare the chains of alpha-carbon atoms calculated with
our algorithm with the corresponding ones obtained from the coordinates
stored at PDB. The resultant rmsd`s are: $0.06$ \AA{} for 2BFI; $%
0.26$ \AA{}\ for 1GCN, $0.14$ \AA{} for 2CK5, $0.29$ \AA{} for 1HG7 and $0.79$ \AA{} for 1MBN.

\begin{figure}[h!]
\centering
\includegraphics[width=0.5\textwidth]{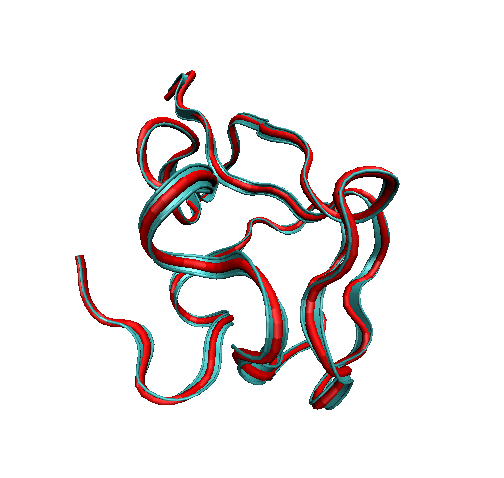} 
\caption{Trace representation of the alpha-carbon atoms backbone for
the protein 1HG7. Red (dark grey) tube: from the coordinates obtained using
our procedure. Cyan (light grey) ribbon: from the coordinates stored at PDB.}
\label{fig:7}
\end{figure}

\begin{figure}[h!]
\centering
\includegraphics[width=0.5\textwidth]{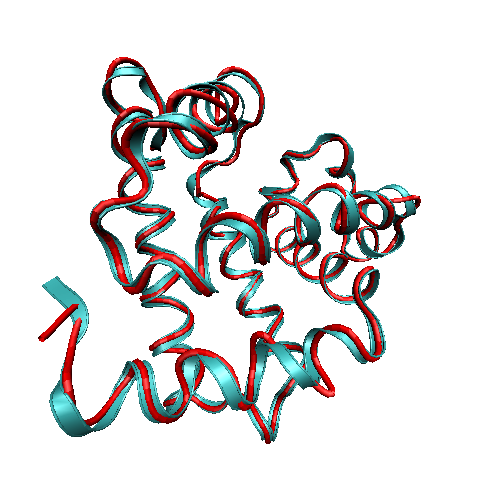} 
\caption{Trace representation of the alpha-carbon atoms backbone for
the protein 1MBN. Red (dark grey) tube: from the coordinates obtained using
our procedure. Cyan (light grey) ribbon: from the coordinates stored at PDB.}
\label{fig:8}
\end{figure}

For 2BFI, 1GCN and 1HG7 we have reconstructed the full-atom protein models
from their alpha-carbon atoms representations using Rotkiewicz and Skolnick
algorithm (PULCHRA)\cite{Rotkiewicz1}. The results are shown in figures 9,
10 and 11 were we also display the corresponding proteins as obtained from
the PDB coordinates.

\begin{figure}[h!]
\centering
\includegraphics[width=0.5\textwidth]{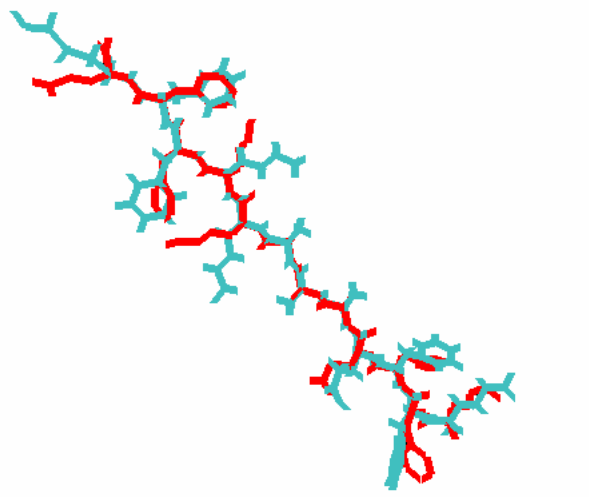} 
\caption{Full atom line representation of the peptide 2BFI. Red (dark
grey): reconstruction from the alpha-carbon atoms backbone coordinates
(obtained with our procedure) using the method of Ref. \cite{Rotkiewicz1}.
Cyan (light grey): from the coordinates stored at PDB. In the rebuilt
protein the hydrogen atoms do not appear.}
\label{fig:9}
\end{figure}

\begin{figure}[h!]
\centering
\includegraphics[width=0.5\textwidth]{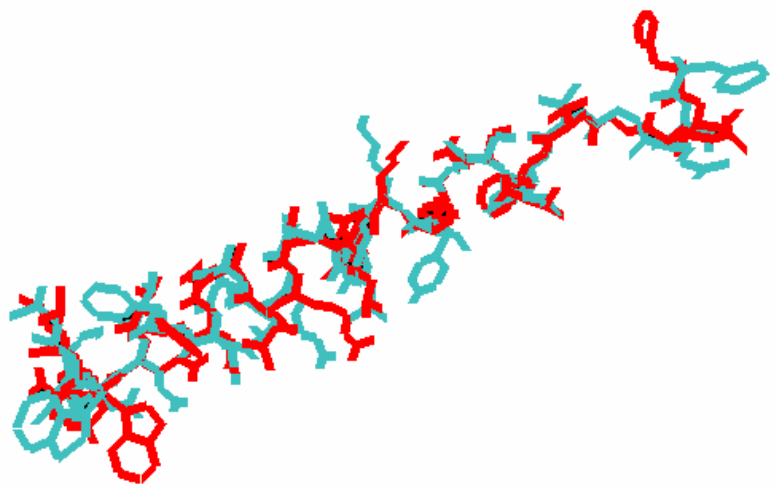} 
\caption{Full atom line representation of the protein 1GCN. Red
(dark grey): reconstruction from the alpha-carbon atoms backbone coordinates
(obtained with our procedure) using the method of Ref. \cite{Rotkiewicz1}.
Cyan (light grey): from the coordinates stored at PDB. In the rebuilt
protein the hydrogen atoms do not appear.}
\label{fig:10}
\end{figure}

\begin{figure}[h!]
\centering
\includegraphics[width=0.5\textwidth]{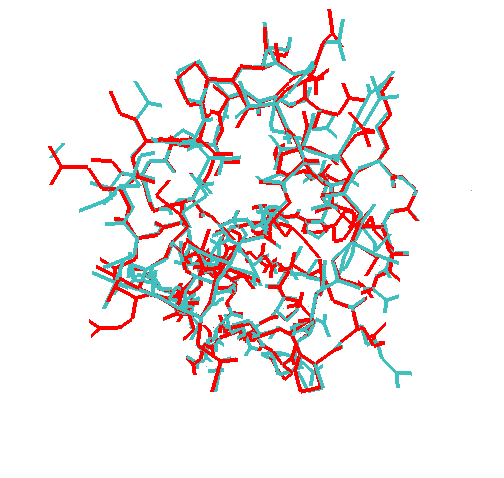} 
\caption{Full atom line representation of the protein 1HG7. Red
(dark grey): reconstruction from the alpha-carbon atoms backbone coordinates
(obtained with our procedure) using the method of Ref. \cite{Rotkiewicz1}.
Cyan (light grey): from the coordinates stored at PDB. In the rebuilt
protein the hydrogen atoms do not appear.}
\label{fig:11}
\end{figure}

It must be remarked again that in this work we simply have shown a way to
pass from the primary to the tertiary structure of the proteins assuming as
known the corresponding order quaternions. These quaternions were obtained
by fitting the coordinates of the alpha-carbon atoms obtained following our
algorithm with the corresponding ones stored at PDB. The problem of using
the procedure here described in order \textit{to} \textit{predict }the
tertiary structure of proteins just from their amino acids sequences, which
implies to know \textit{a priori }a unique set of order quaternions that be
adequate for all proteins, is left for future studies. Despite this
important question, we believe that the results we have obtained until now
already give a good idea about the usefulness of associating amino acids
with quaternions, this being the main objective of this Section.

\section{Conclusions}

In this work we have presented a mathematical representation of the standard
genetic code. Starting from a set of four prime integer quaternions (one for
each of the nucleotide bases that form the mRNA molecules) and guided by a
heuristic diagram that we propose for the evolution of the code, we
introduce a function that assigns an integer quaternion (type quaternion) to
each codon (represented by a triplet of the prime integer quaternions) and
preserves the main properties of the genetic code. The diagram we introduce
for describing the evolution of the genetic code is based in pioneering
ideas by Crick and incorporates, in a way that resembles the energy levels
of an atom, the physical notion of broken symmetry. The objects that we use
for performing the mathematical representation of the code, the Hamilton
quaternions, have as remarkable properties the fact that they verify a non
commutative algebra and their capability for describing spatial rotations.
In particular, this last property gives a special character to the
representation in the sense that it allows to develop a procedure for going
from the primary to the tertiary structure of proteins. To this effect we
introduce a set of real quaternions (order quaternions) that, together with
the integer type quaternions, univocally identify each amino acid of the
proteins. Given an amino acids sequence we present an algorithm that
determines the coordinates of the alpha-carbon atoms of the corresponding
protein using the type and order quaternions. However here we simply adjust
the order quaternions in order to reproduce the experimental coordinates
stored at PDB. As already was commented above, we postpone for future
studies the question of searching for a set of order quaternions which be
common to all the proteins, say the possibility of approaching the protein
folding problem by using our procedure. In our criterion this possibility
distinguishes the above quaternionic representation of the genetic code
among the diverse reported mathematical representations.

\textbf{Acknowledgments}

Support of this work by Universidad Nacional de La Plata, Universidad
Nacional de Rosario and Consejo Nacional de Investigaciones Cient\'{\i}ficas
y T\'{e}cnicas of Argentina is greatly appreciated. The authors are members
of CONICET.

\newpage

\bigskip\bigskip

\begin{center}
\textbf{Appendix A: One and three letters convention for the 20 standard
amino acids}

\bigskip\bigskip

$%
\begin{array}{ccc}
\text{\textbf{Amino acid \ \ }} & \text{\textbf{Three letter}} & \text{%
\textbf{One letter}} \\ 
\text{alanine \ \ \ \ \ \ \ \ \ \ } & \text{ala} & \text{A} \\ 
\text{arginine \ \ \ \ \ \ \ \ \ } & \text{arg} & \text{R} \\ 
\text{asparagine \ \ \ \ \ } & \text{asn} & \text{N} \\ 
\text{aspartic acid \ } & \text{asp} & \text{D} \\ 
\text{cysteine \ \ \ \ \ \ \ \ } & \text{cys} & \text{C} \\ 
\text{glutamic acid } & \text{glu} & \text{E} \\ 
\text{glutamine \ \ \ \ \ } & \text{gln} & \text{Q} \\ 
\text{glycine \ \ \ \ \ \ \ \ \ \ } & \text{gly} & \text{G} \\ 
\text{histidine \ \ \ \ \ \ \ \ } & \text{his} & \text{H} \\ 
\text{isoleucine \ \ \ \ \ \ } & \text{ile} & \text{I} \\ 
\text{leucine \ \ \ \ \ \ \ \ \ \ } & \text{leu} & \text{L} \\ 
\text{lysine \ \ \ \ \ \ \ \ \ \ \ } & \text{lys} & \text{K} \\ 
\text{methionine \ \ \ \ } & \text{met} & \text{M} \\ 
\text{phenylalanine } & \text{phe} & \text{F} \\ 
\text{proline \ \ \ \ \ \ \ \ \ \ } & \text{pro} & \text{P} \\ 
\text{serine \ \ \ \ \ \ \ \ \ \ \ } & \text{ser} & \text{S} \\ 
\text{threonine \ \ \ \ \ \ } & \text{thr} & \text{T} \\ 
\text{tryptophan \ \ \ } & \text{trp} & \text{W} \\ 
\text{tyrosine \ \ \ \ \ \ \ \ } & \text{tyr} & \text{Y} \\ 
\text{valine \ \ \ \ \ \ \ \ \ \ \ } & \text{val} & \text{V}%
\end{array}
$

\newpage

\bigskip

\textbf{Appendix B: Hamilton quaternions}
\end{center}

Quaternions were invented by mathematician William Rowlan Hamilton\cite{Hamilton1,Hamilton2} in 1843 as a generalization of the complex
numbers with the aim of describing rotations in the space in the same sense
as complex numbers describe rotations in the plane. Here we give for
completeness some of the main properties of quaternions. We concentrate
ourselves into their definition, the algebra they fulfill and their relation
with rotations in the space\cite{Kuipers1}.

\bigskip

\textit{Definition}

A quaternion $q$ is an ordered list of four numbers: $q=\left(
a_{0},a_{1},a_{2},a_{3}\right) $ with $a_{0},a_{1},a_{2},a_{3}\in%
\mathbb{R}
$. In the particular case in that the four numbers are integers we talk of
integer quaternions (Lipschitz quaternions). Alternatively we can introduce
the placeholders $\mathbf{i}$, $\mathbf{j}$, $\mathbf{k}$ and represent the
same quaternion as $q=a_{0}+a_{1}\mathbf{i}+a_{2}\mathbf{j}+a_{3}\mathbf{k}$%
. The placeholders $\mathbf{i}$, $\mathbf{j}$, $\mathbf{k}$ verify the
product rules

\begin{align*}
\mathbf{ii} & \mathbf{=-}1\text{ \ \ \ \ \ \ \ \ }\mathbf{jj=-}1\text{ \ \ \
\ \ \ \ }\mathbf{kk=-}1 \\
\mathbf{ij} & \mathbf{=k}\text{ \ \ \ \ \ \ \ \ \ \ }\mathbf{jk=i}\text{ \ \
\ \ \ \ \ \ \ \ }\mathbf{ki=j} \\
\mathbf{ji} & \mathbf{=}-\mathbf{k}\text{ \ \ \ \ \ \ \ }\mathbf{kj=}-%
\mathbf{i}\text{ \ \ \ \ \ \ \ }\mathbf{ik=-j}
\end{align*}
Note that the placeholders play for quaternions a role in some sense similar
to that of the imaginary unit $i=\sqrt{-1}$ for the complex numbers. In this
context the triplet $\left( a_{1},a_{2},a_{3}\right) $ would be the
"imaginary" part of the quaternion. Defining the (real and imaginary)
quaternions $q_{R}=\left( a_{0},0,0,0\right) $ and $q_{I}=\left(
0,a_{1},a_{2},a_{3}\right) ,$we can write: $q=q_{R}+q_{I}$.

\bigskip

\textit{Algebra}

Let $s$ be a real number and $q=\left( a_{0},a_{1},a_{2},a_{3}\right) $, $%
p=\left( b_{0},b_{1},b_{2},b_{3}\right) $ and $r=\left(
c_{0},c_{1},c_{2},c_{3}\right) $ quaternions, we give here the definition of
a few operations:

\medskip

- Conjugation: $\widetilde{q}=\left( a_{0},-a_{1},-a_{2},-a_{3}\right) $

- Scalar multiplication: $sq=\left( sa_{0},sa_{1},sa_{2},sa_{3}\right) $

- Addition of quaternions: $q+p=\left(
a_{0}+b_{0},a_{1}+b_{1},a_{2}+b_{2},a_{3}+b_{3}\right) $

- Multiplication of quaternions: $qp=r$ where

\begin{align*}
c_{0} & =a_{0}b_{0}-a_{1}b_{1}-a_{2}b_{2}-a_{3}b_{3} \\
c_{1} & =a_{0}b_{1}+a_{1}b_{0}+a_{2}b_{3}-a_{3}b_{2} \\
c_{2} & =a_{0}b_{2}-a_{1}b_{3}+a_{2}b_{0}+a_{3}b_{1} \\
c_{3} & =a_{0}b_{3}+a_{1}b_{2}-a_{2}b_{1}+a_{3}b_{0}
\end{align*}
\qquad\ 

\ \ \ \ \ Note that this product is not commutative say, in general, $qp\neq
pq$.

- Norm: $N\left( q\right) =q\widetilde{q}=\widetilde{q}%
q=a_{0}^{2}+a_{1}^{2}+a_{2}^{2}+a_{3}^{2}$

\ \ \ \ \ \ A quaternion $q$ with $N\left( q\right) =1$ is called a unit
quaternion. \ 

\ \ \ \ \ An important property of the norm is that it is multiplicative: $%
N\left( pq\right) =N\left( p\right) N\left( q\right) $\ 

- Inverse: $q^{-1}=\widetilde{q}/N\left( q\right) $ \ \ $(q\neq\left(
0,0,0,0\right) )$

\bigskip

\textit{Quaternions and 3D rotations}

If $N\left( q\right) =1$ then the matrix

\begin{equation*}
R_{q}=\left( 
\begin{array}{cccc}
a_{0}^{2}+a_{1}^{2}+a_{2}^{2}+a_{3}^{2} & 0 & 0 & 0 \\ 
0 & a_{0}^{2}+a_{1}^{2}-a_{2}^{2}-a_{3}^{2} & 2a_{1}a_{2}-2a_{0}a_{3} & 
2a_{1}a_{3}+2a_{0}a_{2} \\ 
0 & 2a_{1}a_{2}+2a_{0}a_{3} & a_{0}^{2}-a_{1}^{2}+a_{2}^{2}-a_{3}^{2} & 
2a_{2}a_{3}-2a_{0}a_{1} \\ 
0 & 2a_{1}a_{3}-2a_{0}a_{2} & 2a_{2}a_{3}+2a_{0}a_{1} & 
a_{0}^{2}-a_{1}^{2}-a_{2}^{2}+a_{3}^{2}%
\end{array}
\right)
\end{equation*}
is a rotation matrix. The oriented axis of rotation $\overrightarrow{e}$ is
given by

\begin{equation*}
\overrightarrow{e}=\frac{\overrightarrow{q}}{\left\vert \overrightarrow {q}%
\right\vert },
\end{equation*}
with $\overrightarrow{q}=a_{1}\widehat{e}_{x}+a_{2}\widehat{e}_{y}+a_{3}%
\widehat{e}_{y}$ where $\widehat{e}_{x}$, $\widehat{e}_{y}$ and $\widehat{e}%
_{z}$ are versors along the three Cartesian axis. The angle $\theta$ that
determines the rotation around the axis $\overrightarrow{e}$ satisfies the
following equation:

\begin{equation*}
\tan\left( \theta/2\right) =\frac{\sqrt{a_{1}^{2}+a_{2}^{2}+a_{3}^{2}}}{a_{0}%
}.\text{ \ \ \ \ \ }
\end{equation*}
Moreover, if we denote with $R_{3\text{ }}$the $3\times3$ matrix that
results when in matrix $R_{q}$ the first row and the first column are
deleted, then we can see that the quaternion $q$ transforms by rotation a
vector $\overrightarrow{r}_{0}=x_{0}\widehat{e}_{x}+y_{0}\widehat{e}%
_{y}+z_{0}\widehat{e}_{z}$ into the vector $\overrightarrow{r}_{1}=x_{1}%
\widehat{e}_{x}+y_{1}\widehat{e}_{y}+z_{1}\widehat{e}_{z}$ according with

\begin{equation*}
\left[ 
\begin{array}{c}
x_{1} \\ 
y_{1} \\ 
z_{1}%
\end{array}
\right] =R_{3\text{ }}\left[ 
\begin{array}{c}
x_{0} \\ 
y_{0} \\ 
z_{0}%
\end{array}
\right] .
\end{equation*}

\end{document}